\documentclass{50}

\def\Lam{\Lambda}
\def\Del{\Delta}

\begin{document}

\title{Quark recombination model for polarizations \\
in inclusive hyperon productions at high energy}

\author{N. Nakajima, K. Suzuki, H. Toki}

\address{Research Center for Nuclear Physics (RCNP), 
Osaka University, \\ Mihogaoka 10-1, Ibaraki, Osaka 567-0047, Japan }

\author{ K.-I. Kubo}

\address{Department of Physics, Tokyo Metropolitan University, \\
Hachiohji, Tokyo 192-0397, Japan}  


\maketitle

\abstracts{
We investigate the transverse polarization of hyperons produced by the photon 
induced reactions using the quark recombination model.
This model reproduces polarizations of hadrons produced by the hadron-hadron 
collisions and accounts for the origin of the empirical rule by 
DeGrand and Miettinen.  
We find 
significant polarizations in the hyperon photoproduction
by applying this model to 
the $\gamma N \ra \Lam $ $(\Sigma^{0}) + X$ reaction.  
}

\section{Introduction}

Significant transverse polarizations and analyzing powers 
of inclusively produced hadrons $\Lambda$
in unpolarized hadron-hadron collisions are observed 
at high Feynman $x_{F}$$(= 2p_L / \sqrt{s}$, $p_L$ is the longitudinal 
momentum of the observed particle) and 
low transverse momentum $p_{T}$ region\cite{bet} against predictions of 
the perturbative QCD\cite{kpr}. 
These data indicate some non-perturbative mechanism being intact for the 
production of the polarization.
Several models are proposed to account for these 
phenomena\cite{dm,ykt,sntk,agl,bzt}.  
It is generally expected that the spin polarization is generated by 
some soft processes in the initial state and/or final state thorough the 
hadron-hadron collision.

Yamamoto {\em et al.}~have constructed the quark recombination model 
to explain the generation of the polarization\cite{ykt,sntk}.  
In this model, 
a final state hadron is produced by the simple recombination of 
relativistic quarks and/or diquarks, and 
polarizations are generated by the scalar 
confining force through the hadronization process in purely 
non-perturbative way.  
It was shown that this model reproduces the empirical rule of
DeGrand and Miettinen(DM)\cite{dm}, and 
provides the polarizations in agreement with 
the experimental data.  
In the framework of this model, we naturally expect non-vanishing 
polarizations even in the hyperon photoproduction.  
Here, we 
investigate polarizations of the hyperons in the photon 
induced reaction using the quark recombination (QRC) model~\cite{dm,ykt}.

\section{Quark Recombination Model}

Let us briefly introduce the quark recombination model\cite{ykt}.  
It is assumed that a fast valence quark (or diquark) directly coming from 
the projectile hadron picks up slow quarks created by the hadronization 
to form the finale state hadron.  The particle production at high $x_{F}$ 
is dominated by this direct process, while the middle and small $x_{F}$ 
hadron production are described by the standard fragmentation, 
in which all the quarks are randomly created by the string breaking.  
Hence, the produced hadron at the high $x_{F}$ contains some information on 
the valence quark structure of the beam hadron.  
In fact, the polarization of the hyperon in the final state reflects the 
spin structure of the beam hadron (see assumption [{\it 4}]).  

Our model is based on the following assumptions;
%
\begin{list}{}
{
\setlength{\topsep}{3pt}%
\setlength{\itemsep}{-1pt}%
\setlength{\leftmargin}{35pt}
\setlength{\rightmargin}{30pt}}%
    \item[\it 1.]  The produced hadron is formed by direct recombination 
    process, since significant polarizations are observed in
    the high $x_{F}$ region.

    \item[\it 2.]  Each parton which participates in this reaction has the 
    intrinsic transverse momentum distribution.

    \item[\it 3.]  Quark and diquark are recombined by the scalar confinement 
    interaction in the  hadronization process.

    \item[\it 4.]  SU(6) spin-flavor symmetry is assumed for the 
initial and final state hadron structure.
\end{list}

Now we concentrate on the $\Lambda$ hyperon production in the 
proton-proton collision. 
In this case, the SU(6) spin-flavor symmetry tells us that the 
valence $(ud)^{0}$-diquark 
from the beam proton picks up a slow $s$-quark 
in order to form the final state $\Lambda$.   
The spin of the $\Lambda$ is determined by the $s$-quark spin.  
Thus, the polarization of the $\Lambda$, $P_N (\Lambda)$, 
is defined as
\bea
P_N (\Lambda) \equiv \frac{\sigma(\Lambda \uparrow) - 
\sigma(\Lambda \downarrow)} {\sigma (\Lambda)} = \frac{\sigma(s \uparrow) - 
\sigma(s \downarrow)} {\sigma (\Lambda)}
\eea
where the spin direction is fixed by $p_{beam} \times p_\Lambda$.

The $\Lambda$ production probability in the Infinite Momentum Frame (IMF) 
is given by
\bea
S_{p \ra \Lam}=
\int[dx_{i}dy_{i}dz_{i}/x_{i}]G_{\Lam}(x_{3},x_{4},y_{3},y_{4},z_{3},z_{4})
\hspace*{3cm}
\nonumber \\ 
\times |M(x_{i},y_{i},z_{i})|^{2}f_{s}(x_{2},y_{2},z_{2})G_{(ud)^{0}/p}
(x_{1},y_{1},z_{1})
\Del^{3}\Del^{4}
\eea
where we choose the $x$-axis as the beam direction and $z$ as the 
orientation of the $\Lam$ spin in the final state, $\Del^{3}$ and $\Del^{4}$ 
express the delta functions which correspond to energy-momentum conservation
in this process.
$G_{(ud)^{0}/p}$ is the momentum distribution of the $(ud)$-diquark in 
the projectile proton. We assume the following form;
\bea
G_{(ud)^{0}/p}(x_{1},y_{1},z_{1})=q_{(ud)^{0}/p} (x_{1}) \; 
e^{-(y_{1}^{2}+z_{1}^{2})}
\eea
where the longitudinal momentum distribution 
$q_{(ud)^{0}/p}(x_{1})$ is the $(ud)$-diquark distribution 
function in the  proton taken from Ref.~\cite{ef}.    
On the other hand, the transverse 
$y$ and $z$ momentum distributions are assumed to be the Gaussian form 
with the average transverse momentum being about 400 MeV.  
$f_{s}$ is the momentum distribution of the piked-up $s$-quark;
\bea
f_{s}(x_{2},y_{2},z_{2})=\theta(x_{1}-x_{2})e^{-(y_{2}^{2}+z_{2}^{2})}.
\eea
where we take the Gaussian transverse momentum distribution.  
We assume that the picked up $s$-quark is slower than the 
$(ud)$-diquark, which is expressed by $\theta$-function.
\vspace{5mm}

\noindent
{
\begin{minipage}[t]{14pc}
\epsfxsize=14pc 
\epsfbox{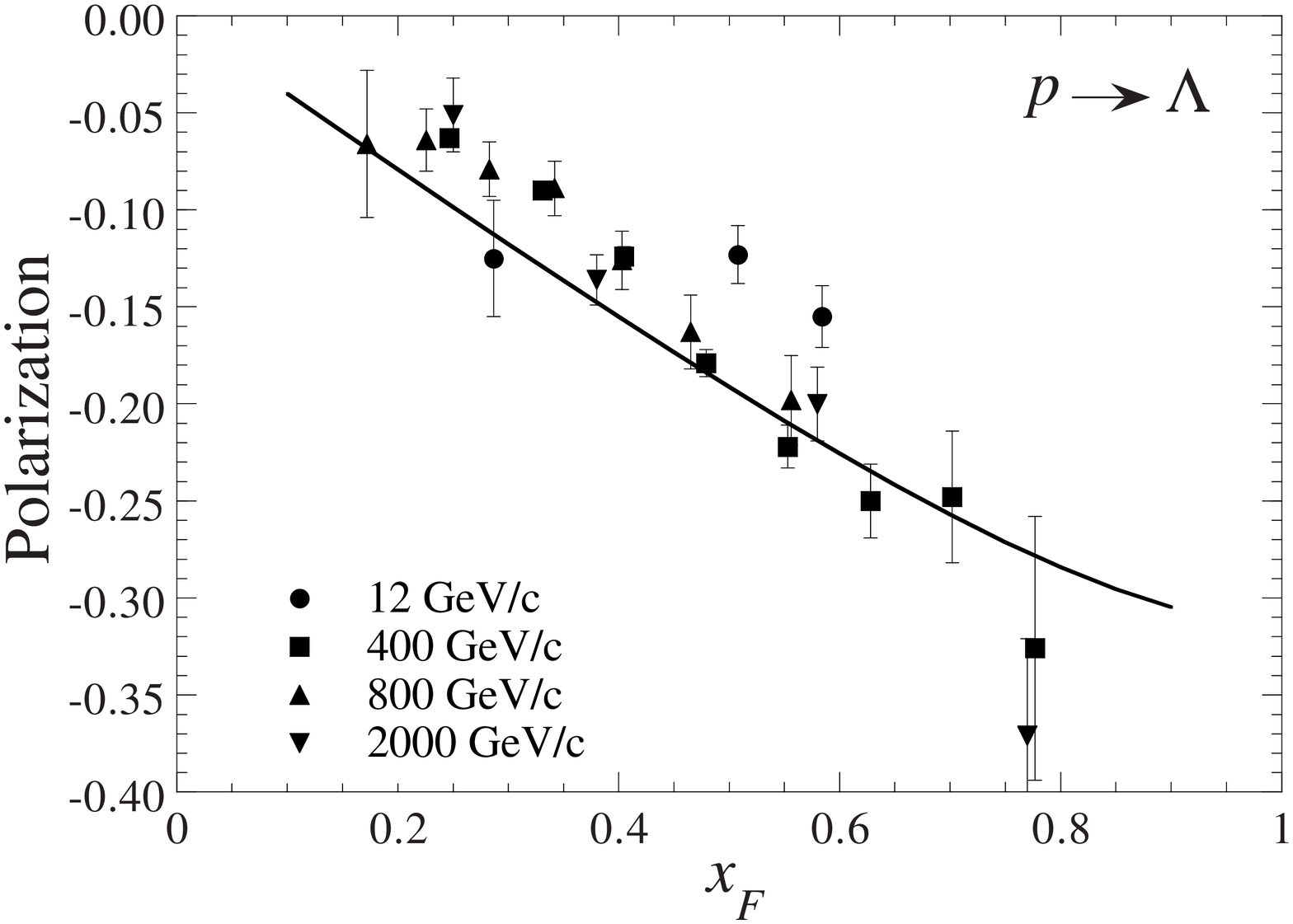} 
\\
{\it Figure 1.} $\Lam$ polarization in the $pp$ collision at 
$p_{T}=1$GeV/$c$ with the experimental data. 
\end{minipage}
}\hfill
{
\begin{minipage}[t]{14pc}
\epsfxsize=14pc 
\epsfbox{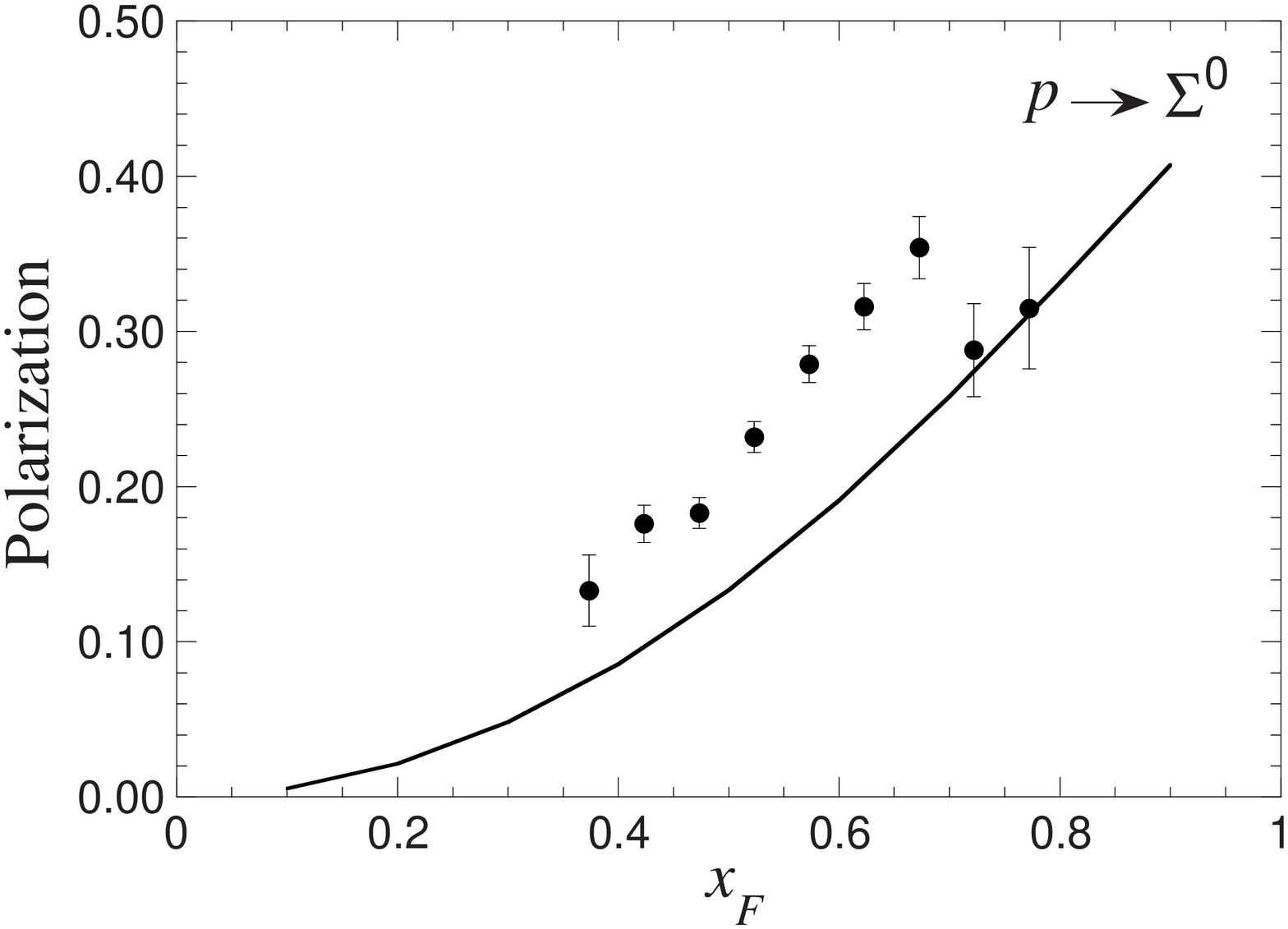} 
\\
{\it Figure 2.} $\Sigma$ polarization in the $pp$ collision 
at $p_{T}=1$GeV/$c$ with the  experimental data.  
\end{minipage}
}
\vspace*{8mm}

\noindent
$G_{\Lam}$ is the light-cone wave function of the final state $\Lam$;
\bea
G_{\Lam}(x_{3},x_{4},y_{3},y_{4},z_{3},z_{4})=
A \exp\left[ -\frac{1}{8\beta^{2}}\left[\frac{k_{t}^{2}+m_{ud}^{2}}{x_{3}}+
\frac{k_{t}^{2}+m_{s}^{2}}{x_{4}}\right]\right].
\eea
$M$ is the elementary hadronization amplitude of the $(ud)$-diquark and 
$s$-quark to produce the $\Lam$ hyperon.  
We shall calculate them explicitly using the scalar confinement 
force, and find that the non-trivial spin dependence arises from the 
interference between the leading and higher order amplitudes.

We finally arrive at the following expression for the polarization,
\bea
P_{N}(\Lam)=R_{0}
\frac
{\int[dx_{i}dy_{i}dz_{i}/x_{i}]G_{\Lam}^{2}\sigma_{dep.}f_{s}G_{(ud)^{0}/p}
\Del^{3}\Del^{4}}
{\int[dx_{i}dy_{i}dz_{i}/x_{i}]G_{\Lam}^{2}\sigma_{ind.}f_{s}G_{(ud)^{0}/p}
\Del^{3}\Del^{4}}
\label{eq:pol1}
\eea
where the spin independent cross section is 
\bea
\sigma_{ind.}=(x_{F}x_{4}x_{2})\left[ 
\left(\frac{x_{F}x_{4}+x_{2}}{x_{F}x_{4}x_{2}}m_{2}\right)^{2}+
\left(\frac{x_{F}x_{4}y_{2}-x_{2}y_{4}}{x_{F}x_{4}x_{2}}\bar{p}_{t}\right)^{2}
\right]
\eea
and the spin dependent one 
\bea
\sigma_{dep.}=-(x_{F}x_{4}x_{2})
\left(\frac{x_{F}x_{4}y_{2}-x_{2}y_{4}}{x_{F}x_{4}x_{2}}\bar{p}_{t}\right).
\eea
Note that, if we took the vector interaction instead of the scalar, the spin 
dependent part of the cross section would vanish in the IMF.  
$R_{0}$ in Eq.~(\ref{eq:pol1}) is a free parameter and will be 
fixed to reproduce the experimental data of the $\Lam$ polarization.

Similarly, we can obtain the formula for the polarization of  
$\Sigma^{0}$ production.  Since the $\Sigma$ hyperon has the 
isospin 1, $\Sigma^{0}$ hyperon is  
formed by the recombination of a fast spin-$1$ $ud$-diquark 
and a slow $s$-quark.  
We need another parameter $R_{1}$ to be fixed by the $\Sigma$ polarization 
data.  
From experimental data for $\Lam$ and $\Sigma$, we determine 
$R_{0}=-12$ GeV and $R_{1}=50$ GeV.  
Our results are shown in Fig. 1 and 2 for the $\Lambda$ and $\Sigma$ 
polarizations, which are consistent with experiments.  
We can calculate the polarization of other hadrons, e.g.~$\Xi$, which is also 
in 
good agreement with experiments.  

\section{Photon induced hyperon production }

In our model, polarizations are generated by the non-perturbative 
hadronization process.
Therefore, even in the photon induced reaction, we expect 
finite polarizations of the produced hyperons, because the hadronization 
process itself is independent of the beam.  
Let us consider the unpolarized $\gamma$-hadron reaction as schematically 
shown in Fig. 3.  
It is known that the real photon has the hadronic (vector meson) structure, 
since the real 
photon has enough time to turn into a $\bar q q$ system.  
In fact, the quark distribution of the real photon observed by the deep 
inelastic lepton-photon scattering contains the substantial hadronic 
components.
We show in Fig. 4 the parametrization of the quark distribution of the 
real photon taken from ref.~\cite{grv}.  

We apply our model to the photoproduction of hyperons.  
we replace the diquark distribution of the proton in the previous formula 
Eq.~(\ref{eq:pol1}) with the quark distribution of the photon $G_{q/\gamma}$. 
In this case, the $u$(or $d$ or $s$)-quark picks up a diquark to 
form the $\Lam$ or $\Sigma^{0}$ hyperons.  
Notice that both spin-$0$ and spin-$1$ diquark recombination processes 
contribute to this reaction.
We get the following expression for the $\Lam$ polarization in the photon 
induced reaction,
\bea
P_{N}(\Lam)=
\frac
{\sum_{l} R_{l}\int[dx_{i}dy_{i}dz_{i}/x_{i}]G_{\Lam}^{2}
\sigma_{dep.,l}f_{(qq)^{l}}G_{q/\gamma} \Del^{3}\Del^{4}}
{\sum_{l} \int[dx_{i}dy_{i}dz_{i}/x_{i}]G_{\Lam}^{2}
\sigma_{ind.,l}f_{(qq)^{l}}G_{q/\gamma} \Del^{3}\Del^{4}}
\label{eq:pol2}
\eea
where $l$ runs over possible combinations of quarks and diquarks.  

We show in Fig. 5 and 6 our results for the $\Lam$ and $\Sigma$ polarizations 
in the $\gamma N$ reaction.  
The magnitude of the polarization is similar with that of the proton-proton 
collision.  These theoretical predictions will be clarified by future 
experiments such as the PEARL facility.  
\vspace{10mm}

\noindent
{
\begin{minipage}[t]{14pc}
\epsfxsize=14pc 
\epsfbox{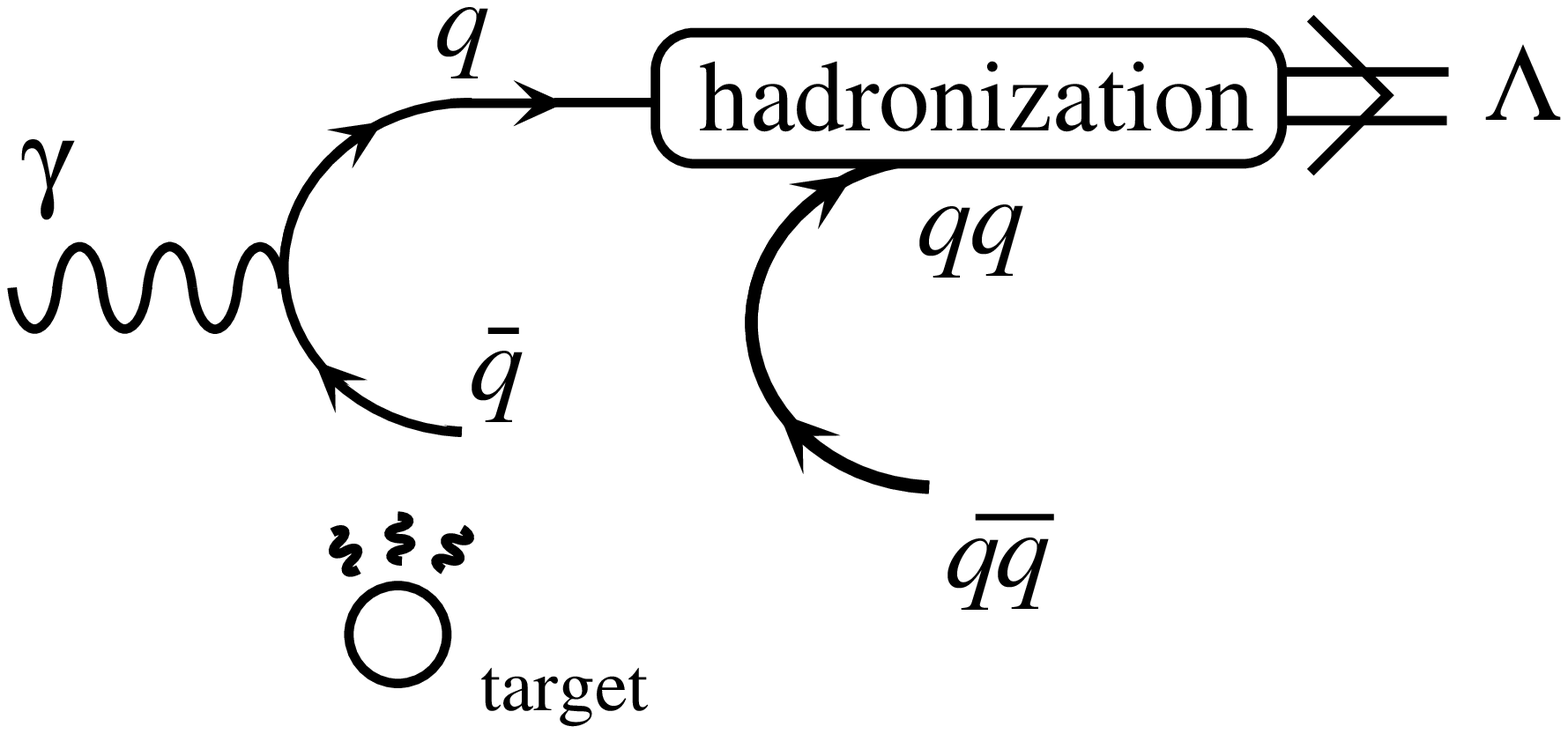} 
\\
{\it Figure 3. Hyperon production by the $\gamma$-hadron reaction } 
\end{minipage}
}
\hfill
{
\begin{minipage}[t]{14pc}
\epsfxsize=14pc 
\epsfbox{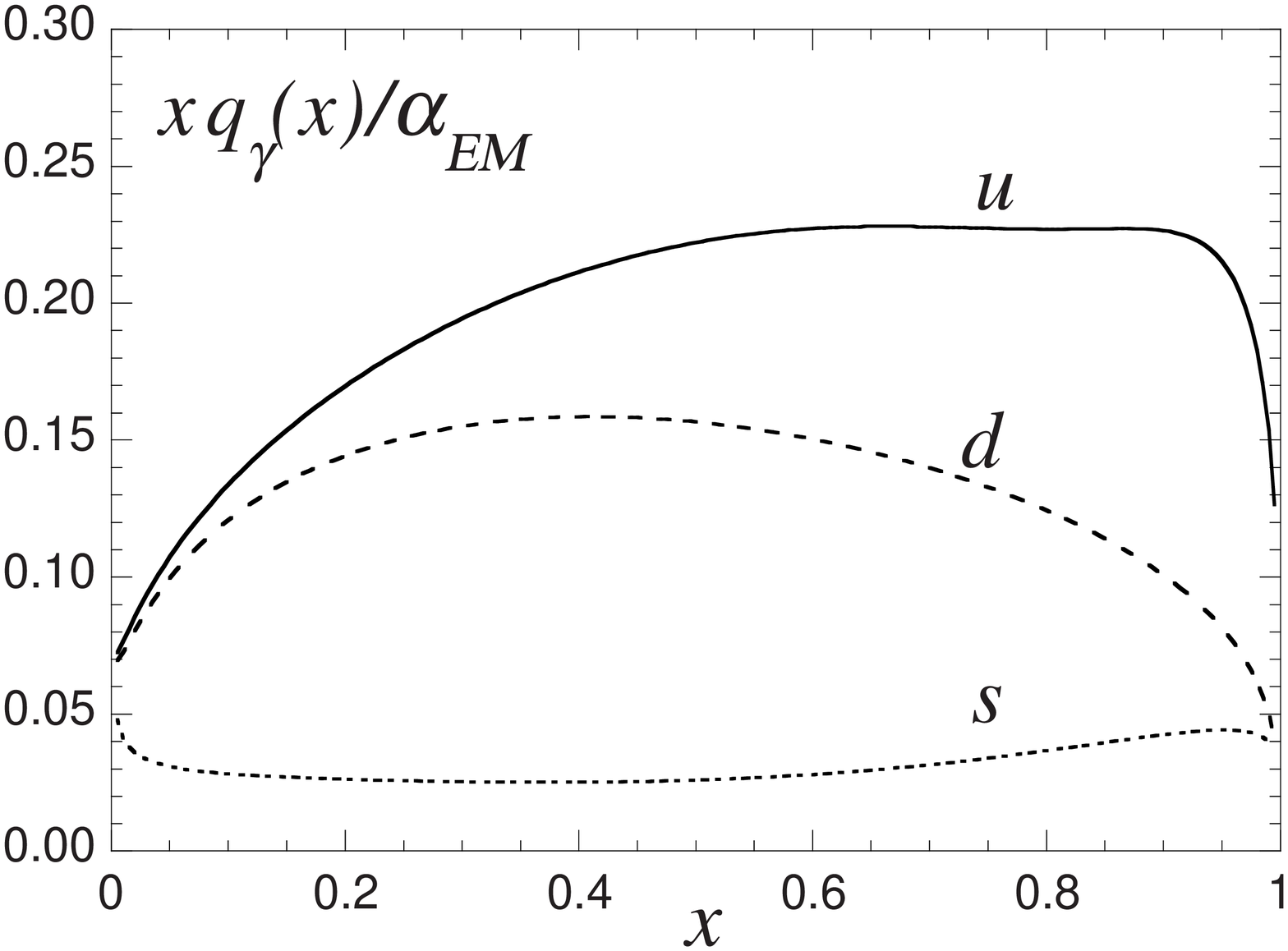} 
\\
{\it Figure 4.} Photon structure function taken from ref.~\cite{grv}
\end{minipage}
}

\vspace{.9cm}

\newpage
\section{Summary}

In conclusion, we have studied the spin polarizations of the hyperon 
photoproduction based on the quark recombination model.  
Our model may be applicable for the process with the photon beam 
, whose energy is larger than about 5GeV, 
where the quark degrees of freedom are essential rather than 
the hadronic components.  
We have reproduced 
various polarizations in the inclusive hyperon production 
in the hadron-hadron collisions 
with the model parameters being fixed to reproduce the absolute 
magnitudes of the polarizations of $\Lambda$ and $\Sigma$.  
With these parameters we have found that about $20 \sim 30 \%$ 
negative polarizations 
of $\Lam$ in the photon induced reaction, which will 
be tested by future experiments. 
It is also possible to extend the present model to the $e^+ e^-$ 
annihilation process.  However, the polarization almost vanishes in the 
$e^+ e^- \to \Lambda +X$ case, because the intrinsic transverse momentum 
distribution of the initial quark can be neglected in this case.  

\vspace{5mm}

{
\begin{minipage}[t]{13pc}
\epsfxsize=14pc 
\epsfbox{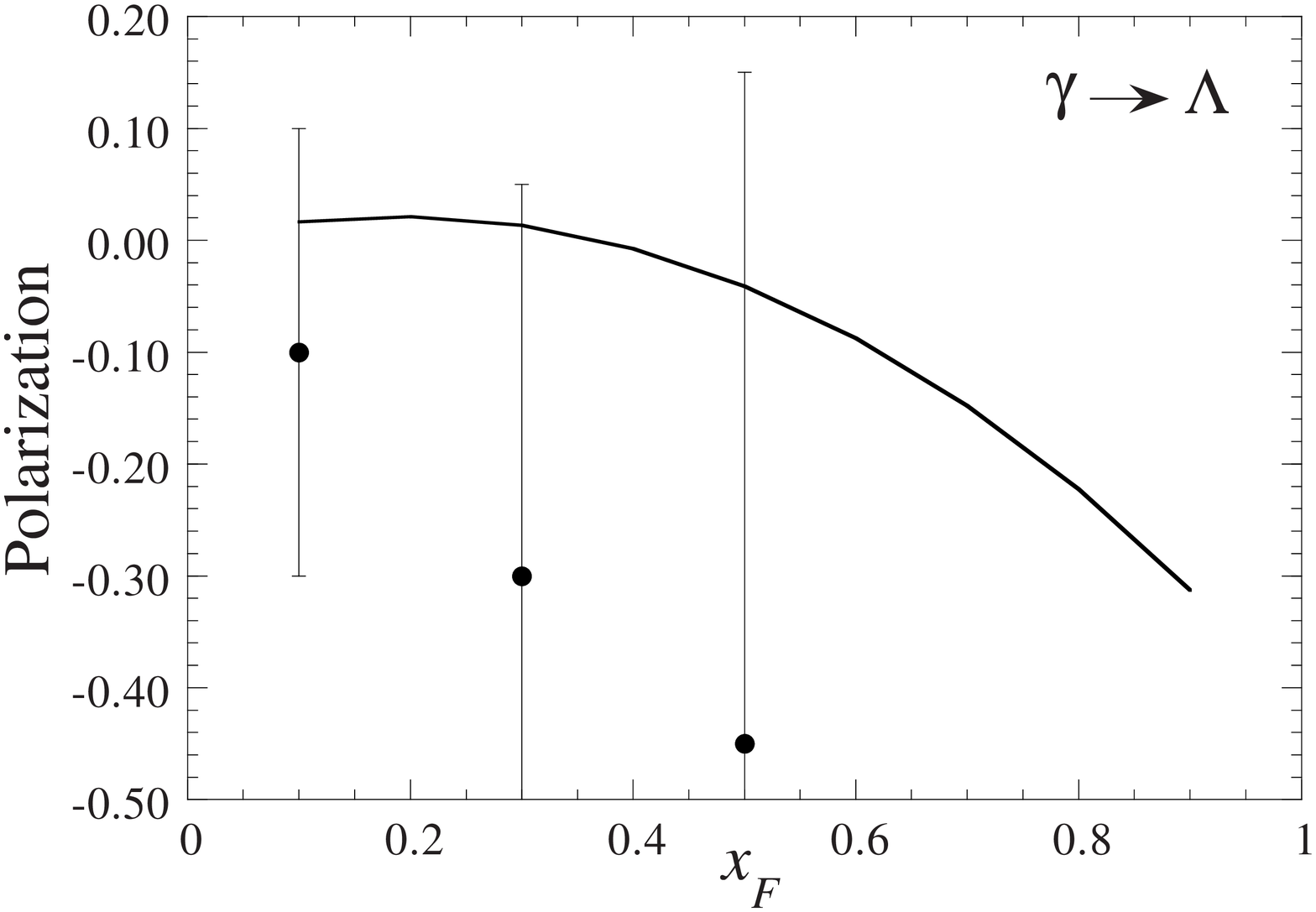} 
\\
Figure 5. $\Lam$ Polarization in the $\gamma N$ reaction.  Experimental 
data are taken from ref.~\cite{abe}.  

\label{fig:pglam}
\end{minipage}
}
\hfill
{
\begin{minipage}[t]{13pc}
\epsfxsize=14pc 
\epsfbox{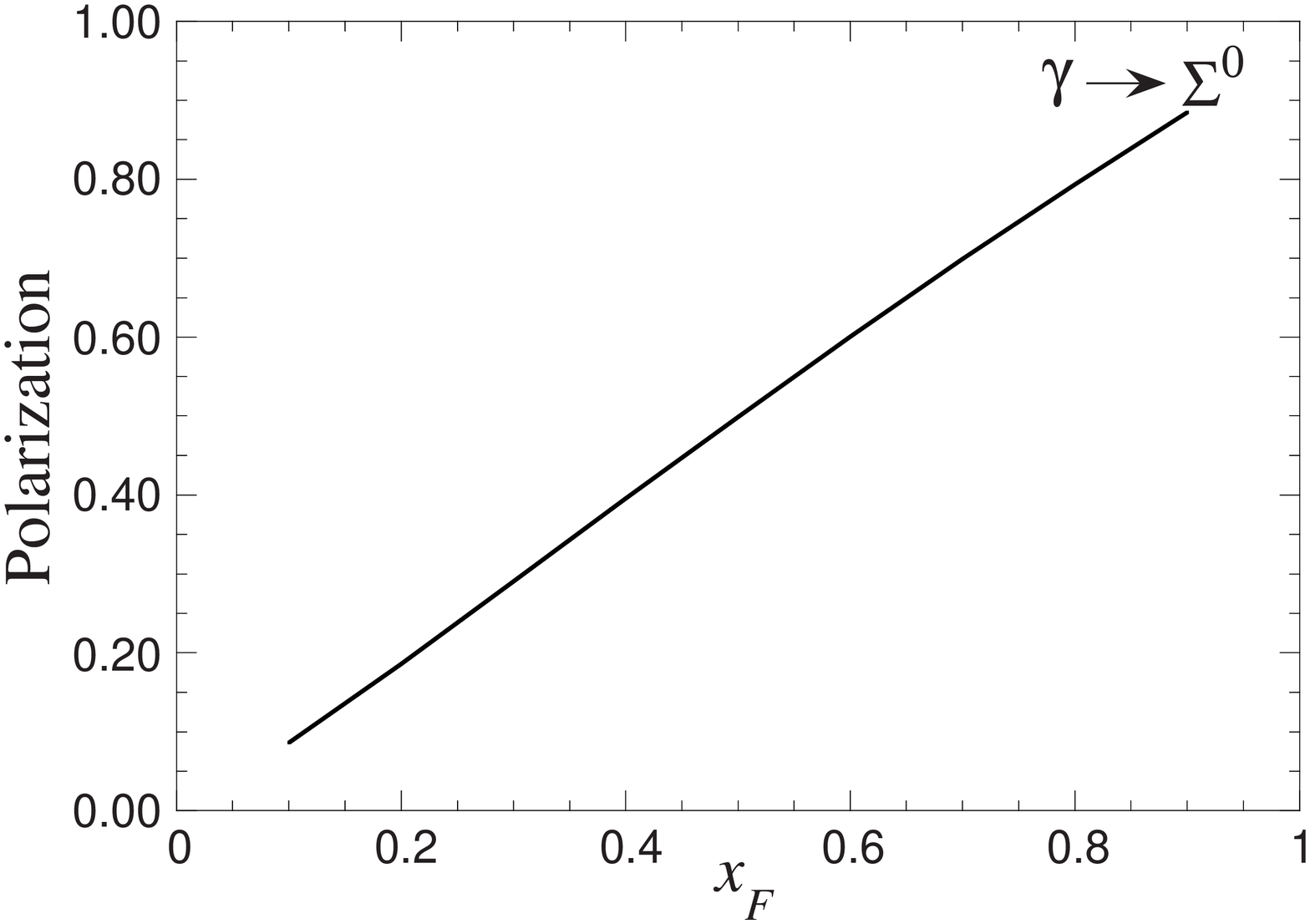} 
\\
Figure 6. $\Sigma$ Polarization in the $\gamma N$ reaction. 
\label{fig:pgsig}
\end{minipage}
}

\end{document}